\documentclass[10pt,a4paper]{amsart}
\usepackage{geometry}               
\usepackage{graphicx}
\usepackage{amssymb}
\usepackage{epstopdf}
\usepackage{url}
\begin{document}
\title{A Potentiality and Conceptuality Interpretation of Quantum Physics}
\author{Diederik Aerts}
\address{Leo Apostel Center for Interdisciplinary Studies, Brussels Free University, Krijgskundestraat 33, 1160 Brussels, Belgium}
\email{diraerts@vub.ac.be}
\urladdr{http://www.vub.ac.be/CLEA/aerts/}
\begin{abstract}
We elaborate on a new interpretation of quantum mechanics which we introduced recently. The main hypothesis of this new interpretation is that quantum particles are entities interacting with matter conceptually, which means that pieces of matter function as interfaces for the conceptual content carried by the quantum particles. We explain how our interpretation was inspired by our earlier analysis of non-locality as non-spatiality and a specific interpretation of quantum potentiality, which we illustrate by means of the example of two interconnected vessels of water. We show by means of this example that philosophical realism is not in contradiction with the recent findings with respect to Leggett's inequalities and their violations. We explain our recent work on using the quantum formalism to model human concepts and their combinations and how this has given rise to the foundational ideas of our new quantum interpretation. We analyze the equivalence of meaning in the realm of human concepts and coherence in the realm of quantum particles, and how the duality of abstract and concrete leads naturally to a Heisenberg uncertainty relation. We illustrate the role played by interference and entanglement and show how the new interpretation explains the problems related to identity and individuality in quantum mechanics. We put forward a possible scenario for the emergence of the reality of macroscopic objects.  
\end{abstract}
\maketitle
\section{Introduction}
It is commonly accepted that the micro-world described by quantum theory is fundamentally different from how we would imagine it to be based on our everyday experience of the macroscopic physical world around. Indeed, textbooks and articles on quantum theory generally tell us that quantum particles are not like minute ping pong balls bumping and bouncing around. But has this fundamental difference between the nature of the micro-world and the nature of the everyday macro-world been fully digested? In this article we will explore possible answers to this key question, discussing our recent proposal for a new interpretation of quantum mechanics \cite{aerts2009a,aerts2010a}.

Although there is general agreement that quantum particles are not like tiny little ping pong balls, the question of what they are has not been given much attention. Most physicists are prepared to accept that quantum particles are very strange objects, if objects at all. A substantial group of physicists even doubt whether quantum particles  are objects `in the sense of what we, or even philosophers, imagine objects to be'. And some go as far as to suggest that we had best abandon any attempts to imagine what type of things quantum particles are. In our recently proposed interpretation of quantum theory, we introduce a well-defined proposal for the nature of a quantum particle, namely that it is not an object but a concept \cite{aerts2009a,aerts2010a}. In the present article, we will elaborate on this new interpretation.

There are various reasons why we believe we should try and look for new interpretations of quantum mechanics with the aim of finding explanations, including for such fundamental questions as `What is a quantum particle?'. We even think that young physicists and philosophers should be encouraged to do so. One of the reasons why we feel that it is feasible to make real progress in our understanding of quantum mechanics, is that we believe human imagination to be quite limited when it comes to `imagining and understanding' situations that require an approach that is fundamentally different from that actually believed to be the true one. History abounds with examples to illustrate this point, one of the famous ones being the Copernican revolution \cite{copernicus1543}. For many reasons, but mainly because of daily observation, there was a deeply rooted belief that the sun turned around the earth, so that it was not obvious at all to `imagine and understand' that the actual state of affairs was quite the contrary, namely the earth turning around the sun. The perspective offered by daily observation was also wrong in a more subtle way, partly originating from the difficulty for people to imagine and understand that the earth turned around the sun while they did not feel it moving. They had to realize that it was possible to be on an object moving through space even though they were not aware of this motion. The rise of classical mechanics, with its elaborate and complex explanatory framework, illustrates the potential depth of scientific theory and enquiry. Of course, we now know that classical mechanics was not the end of the story, having been convincingly superseded by general relativity theory. In turn, the latter theory is not likely to have the final say either, if we consider gravitation's well-known resistance to unification with the other three fundamental forces of nature \cite{mohapatra2003,schumm2004,smolin2006,wilczek2008}.

Human imagination is not only limited if it comes to situations requiring a profound change in perspective to be understood. It is not very powerful in a more superficial way either. If we try to imagine how things happen, even in the ordinary world around us, `using only our mind and imagination', we often fail to come up with a scenario that offers a plausible explanation. Illusionists are aware of this and use this lack of imaginary power of the mind to their advantage. And with respect to certain improbable events in the ordinary world, it is often pointed out `how reality can sometimes turn out to be much more unexpected and unbelievable than we could have imagined'. The SF worlds created by imaginative authors, for instance, tend to become outdated over time, proving far less fascinating than reality itself. One telling example is the World-Wide Web, conspicuously absent from any science-fictional world presented around the middle of the previous century.

I came upon an instance of the latter, more superficial limitation of human imagination in the period of time when I worked on the problem of the violation of Bell inequalities \cite{einsteinpodolskyrosen1935,bohm1952,bell1964,bell1966,clauserhorne1974,clausershimony1978,aspect1981,bell1987}, back in the 1980s. By examining concrete, not very complicated ordinary physical systems in the macroscopical world, more specifically `two water vessels interconnected by a tube', I was able to understand many subtleties of the violation of Bell inequalities that could not be understood at all, or at least not in a clear way, through pure abstract reasoning \cite{aerts1982a,aerts1985,aerts1991,aertsaertsbroekaertgabora2000,aertsczachordhooghe2006}. The lesson I learned from this is that it is very fruitful to try and find ways to model abstract situations on concrete, often almost engineering-like configurations and realizations. The reason why I recall this, is that the same approach of looking for concrete configurations and realizations has certainly also played an important role in the new interpretation of quantum mechanics put forward in \cite{aerts2009a,aerts2010a}, which I intend to illustrate in the present article. It is in this context that I will now briefly return to the example of the water vessel experiment, explaining how it inspired this new interpretation.

\section{Quantum and Potentiality}

We consider two vessels $V_A$ and $V_B$ interconnected by a tube $T$, each of them containing 10 liters of transparent water. Coincidence experiments $A$ and $A'$ consist in siphons $S_A$ and $S_B$ pouring out water from vessels $V_A$ and $V_B$, respectively, and collecting the water in reference vessels $R_A$ and $R_B$, where the volume of collected water is measured, as shown in Figure 1. If more than 10 liters is collected for experiments $A$ or $B$ we put $E(A)=+1$ or $E(B)=+1$, respectively, and if less than 10 liters is collected for experiments $A$ or $B$, we put $E(A)=-1$ or $E(B)=-1$, respectively. We define experiments $A'$ and $B'$, which consist in taking a small spoonful of water out of the left vessel and the right vessel, respectively, and verifying whether the water is transparent. We have $E(A')=+1$ or $E(A')=-1$, depending on whether the water in the left vessel turns out to be transparent or not, and $E(B')=+1$ or $E(B')=-1$ depending on whether the water in the right vessel turns out to be transparent or not. We define $E(AB)=+1$ if $E(A)=+1$ and $E(B)=+1$ or $E(A)=-1$ and $E(B)=-1$, and $E(A,B)=-1$ if $E(A)=+1$ and $E(B)=-1$ or $E(A)=-1$ and $E(B)=+1$, if the coincidence experiment $AB$ is performed. Note that we follow the traditional way of defining the expectation value of the coincidence experiments $AB$. In a similar way, we define $E(A'B)$, $E(AB')$ and $E(A'B')$, the expectation values corresponding to the coincidence experiments $A'B$, $AB'$ and $A'B'$, respectively. 
\begin{figure}[h]
\centerline {\includegraphics[width=11cm]{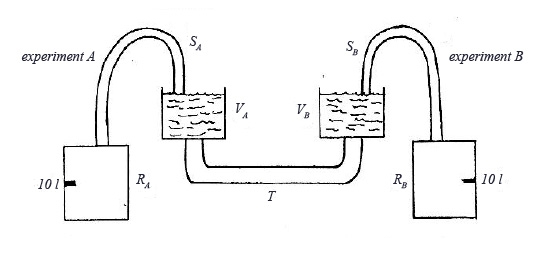}}
\caption{Two vessels $V_A$ and $V_B$ containing 10 liters of transparent water each are interconnected by a tube $T$. Coincidence experiments $A$ and $B$ consist in siphons $S_A$ and $S_B$ pouring out water from vessels $V_A$ and $V_B$, respectively, and collecting the water in reference vessels $R_A$ and $R_B$, where the volume of collected water is measured.}
\end{figure}
\noindent
Hence, concretely $E(A'B)=+1$ if $E(A')=+1$ and $E(B)=+1$ or $E(A')=-1$ and $E(B)=-1$ and the coincidence experiment $A'B$ is performed. And we have $E(AB')=+1$ if $E(A)=+1$ and $E(B')=+1$ or $E(A)=-1$ and $E(B')=-1$ and the coincidence experiment $AB'$ is performed, and further $E(A'B')=+1$ if $E(A')=+1$ and $E(B')=+1$ or $E(A')=-1$ and $E(B')=-1$ and the coincidence experiment $A'B'$ is performed. Since each vessel contains 10 liters of transparent water, we find experimentally that $E(AB)=-1$, $E(A'B)=+1$, $E(AB')=+1$ and $E(A'B')=+1$, which gives
\begin{equation}
E(A',B')+E(A'B)+E(AB')-E(AB)=+4
\end{equation}
This is the maximum possible violation of Bell inequalities. First of all, I should stress that the foregoing is a true representation of a genuine violation of Bell inequalities by means of a macroscopic situation involving only classical macroscopic physical systems. What I mean to say is that there is no misrepresentation involved. The occurrence of this violation of Bell inequalities is also consistent with all theoretical analyses of the Bell inequalities that I am aware of. For example, the `locality condition' used by Bell and others to derive the validity of the inequalities for `local' situations is violated also by this example, as we showed in \cite{aerts1991}. Here is a brief summary. Let us suppose that there is a deterministic hidden variable theory underlying the situation of the experiment, and let us denote by $\Gamma$ the set of these variables $\lambda$. In such a hidden variable description, $E(A,B)$ has a determined value $E(A,B,\lambda)$ for every value $\lambda \in \Gamma$. The locality condition put forward by John Bell is the following \cite{bell1964}. For the measurement situations $AB$, $A'B$, $AB'$ and $A'B'$, and for an arbitrary $\lambda \in \Gamma$ we have
\begin{eqnarray}
E(A,B,\lambda)&=&E(A,\lambda)\cdot E(B,\lambda) \quad E(A',B,\lambda)=E(A',\lambda)\cdot E(B,\lambda) \\
E(A,B',\lambda)&=&E(A,\lambda)\cdot E(B',\lambda) \quad E(A',B',\lambda)=E(A',\lambda)\cdot E(B',\lambda) \nonumber
\end{eqnarray}
If we consider the correlation experiment $AB$, it is easy to specify the hidden variables which make the outcomes determined. For example, if we specify the diameters $\lambda_A$ and $\lambda_B$ of siphons $S_A$ and $S_B$, respectively, the outcomes of the correlation experiment $AB$ are determined. Hence, more specifically, as a simple model, we can write $E(A,\lambda_A,\lambda_B)=+1$ and $E(B,\lambda_A,\lambda_B)=-1$ if $\lambda_B < \lambda_A$, and $E(A,\lambda_A,\lambda_B)=-1$ and $E(B,\lambda_A,\lambda_B)=+1$ if $\lambda_A < \lambda_B$. This is a correct hidden variable description, with corresponding factorization of the expectation value, with respect to the correlation experiment $AB$. However, if we want to use the same $E(A,\lambda_A,\lambda_B)$ to factorize the expectation value $E(A,B')$ for the coincidence measurement $AB'$, it does not work. Indeed, $A$ performed together with $B'$ will always give more than 10 liters of water in vessel $R_A$. This means that the value of $E(A)$ depends not only on the values of $\lambda_A$ and $\lambda_B$ but also on whether we perform $A$ jointly with $B$ or jointly with $B'$, and more specifically
$E(A,\lambda_A,\lambda_B,B) \not= E(A,\lambda_A,\lambda_B, B')$, since $E(A,\lambda_A,\lambda_B,B)=+1$ if $\lambda_B < \lambda_A$ and $E(A,\lambda_A,\lambda_B,B)=-1$ if $\lambda_A < \lambda_B$ while $E(A,\lambda_A,\lambda_B,B') = +1$ for all $\lambda_A, \lambda_B$.

John Bell put forward the locality hypothesis based on the entity consisting of two spin-1/2 particles in the singlet spin state introduced by David Bohm \cite{bohm1952} as an example of the Einstein Podolsky Rosen situation \cite{einsteinpodolskyrosen1935}. Why do most scientists seem to find this locality hypothesis `natural' for this entity? Because they imagine the entity to be an entity consisting of two spin-1/2 particles located in different and widely separated regions of space and flying in opposite directions. And indeed, for two entities located in widely separated regions of space, with no connection between them, the Bell locality hypothesis is a natural hypothesis to be satisfied. But for two entities that actually form a whole, like the water in the two vessels, it is very easy to violate the Bell locality hypothesis, and hence also the Bell inequalities.

Because we can `see with our eyes' what happens in the case of the water vessels, we can understand different aspects of such a type of non-locality situation. For example, one way of looking at the situation would be to say that the four pairs of measurement situations are not compatible, in the sense that if we look for their underlying physical reality, i.e. the reality described in classical deterministic terms, we will find that this underlying physical reality is described differently -- i.e. using other variables -- depending on which measurement couple we start with. If we consider the measurement situation of couple $AB$, the parameters describing the diameter of the two siphons used will play a determining role in the statistics of the outcomes of the correlation measurement. This is not the case for the other three measurement situations $A'B$, $AB'$ and $A'B'$. At the same time, however, since we can easily understand what is happening during the measurements in this example about the interconnected water vessels, we realize that this instance of incompatibility is no obstacle to a realistic interpretation of the vessels of water.

It is also easy to see that the correlations in the case of the joint measurement $AB$ are created `as a consequence and during the act of measurement' and that there is no signal involved going from one side to the other side or vice versa. In \cite{aerts1991}, we called `correlations of the second kind' those correlations that exist only potentially before the measurement, and are actualized as a consequence of the measurement. We believe that the quantum correlations violating the Bell inequalities are correlations of the second kind, much like those present in this example about the water vessels interconnected by a tube. For an analysis of further quantum-like aspects that can be understood and explained using this example, we refer to our articles published on the subject \cite{aerts1982a,aerts1985,aerts1991,aertsaertsbroekaertgabora2000}. 

\section{Realism and Space-Time}

In recent years, an interesting new aspect was put forward by Anthony Leggett, who derived a new set of inequalities with the aim of investigating whether the quantum type of non locality could be of a type which he calls `crypto-non-locality', and he showed that predictions of quantum mechanics violated also these new inequalities \cite{leggett2003}. In the case of the pairs of quantum particles in a singlet spin state, hidden variables are crypto-non-local if it is such that each of the pairs of quantum particles `has' as definite state of spin after the pair has been created in the cascade and before the pair is being detected. Meanwhile, Leggett's inequalities have also been to shown to be violated experimentally, confirming the predictions of quantum mechanics \cite{groblacherpaterekkaltenbaekbruknerzukowskiaspelmeyerzeilinger2007,paterekfedrizzigroblacherjenneweinzukowskiaspelmeyerzeilinger2007,branciardlinggisinkurtsieferlamaslinaresscarani2007,brianciardbrunnergisinkurtsieferlamaslinareslingscarani2008,eisamangoldschmidtchenfanmigdall2008}. While the violation of the Bell inequalities pointed at locality as the problematic hypothesis with respect to quantum mechanics, and hence also with respect to the micro-world, realism was less questioned because it was generally accepted that the loss of locality would be less problematic than the loss of realism. Leggett's analysis, his proposed inequalities and their violations by quantum mechanics and experiments, are taken to indicate that realism is the more problematic hypothesis. This shift towards realism being considered the more problematic hypothesis is felt as a very dramatic one, as is apparent from the fact that in a considerable number of popular writings following the experimental violations of Leggett's inequalities reference was made to Einstein's quote `Does the moon still exist if we do not look at it?'. Indeed, if the problem with realism as a consequence of the violations of Leggett's inequalities means that the moon does not exist if we do not look at it, this is a far more serious problem for understanding quantum mechanics than non-locality could ever be. Our example of the vessels of water shows us how we can reflect in a more nuanced way and arrive at a less dramatic conclusion. We should note that the vessels of water experiment does indeed violate Leggett's inequalities, and it is even meant to do so, since it is an example of the production of what we have called correlations of the second kind, i.e. correlations that do not exist before the measurement takes place, so that they are (partly) created by the measurement. However, does this mean that the experiment causes a problem with respect to philosophical realism? We think not, because, if this were the case, philosophical realism would have encountered similar obstacles long before the advent of quantum mechanics. Events such as those happening in our experiment are abundant in everyday's macroscopic world, and have been so ever since mankind began to form worldviews.

Looking more carefully at the coincidence experiment $AB$, we can see that the water in both vessels and in the tube forms one whole before the correlation experiment takes place. It is the experiment itself that divides the water into two parts, one part flowing to the left, and collected in reference vessel $R_A$, and the other supplementary part flowing to the right, and collected in reference vessel $R_B$. Consequently, after the measurement the water is divided into two parts, and, although the division depends on the details of each individual measurement, the volumes of both parts will always add up to 10 liters. Suppose we denote the different states of the totality of the water after the correlation measurement as follows $p_A(x)\otimes p_B(10-x)$, where $x$ is the amount of water in reference vessel $R_A$ after the coincidence experiment $AB$, and suppose that, for the sake of simplicity, we measure $x$ only in units of liters. This means that we have a total number of 11 states which we distinguish as final states after the measurement $AB$, which are the following $\{p_A(x)\otimes p_B(10-x)\ \vert x\in\{0,1,\ldots,10\}\}$. It is possible to determine such intervals for the values of the diameters of the two siphons that if the values of both diameters are contained in such a product of two intervals, the situation will evolve towards the corresponding end state, and the corresponding outcome occurs. In other words, the description of this one measurement situation $AB$ is completely deterministic. What is now the state of the two vessels of water connected by a tube? All the final states with respect to the coincidence measurement $AB$ are only `potential' states of the two vessels. We could also say that they are `created' by the measurement situation. Of course, they are not created `out of nothing': the state of the vessels of water connected by a tube `in the absence of any measuring' does play a crucial role in `how they are created'. In fact, it is the interplay between the measurements and this state which defines the final states and the outcomes. We believe that it is correct to say that this `pre-measurement state' is a superposition state of these final states, so that we can write it as follows
\begin{equation}
p=\sum_{x=1}^{11}\lambda(x) p_A(x)\otimes p_B(10-x)
\end{equation}
To bring this description as close as possible to a quantum description, we can choose $\lambda(x)$ to be complex numbers, such that $|\lambda(x)|^2$ corresponds to the probability that the final state $p_A(x)\otimes p_B(10-x)$, and outcome $x$ liters in vessel $R_A$ and $10-x$ liters in vessel $R_B$ is reached, which means that we also have $\sum_{x=1}^{11}|\lambda(x)|^2=1$. Obviously, technically speaking, $p$ is an entangled state.

At this point, we should first make clear what we think is quantum-like in this example, and what is not. We do believe that the quantum entangled (or more generally, superposition) states of quantum particles are effectively of this type. They are genuinely different from the final states of the quantum particle after measurements. And they are such that `these final states are only potentially present in the quantum superposition state', in exactly the same way as the 10 liters of water are only potentially subdivided into $x$ liters and $10-x$ liters. In all, we believe that the example of the vessels of water grasps in a correct way this aspect of quantum mechanics, at the same time showing that essentially it is not so mysterious for this state of affairs to occur in nature -- water is an everyday example. Furthermore, it is easy to see that this is not related essentially to the substance `water', but that it is related to `one whole of material being potentially divided into parts', depending on `how the measurement bringing about this division works out in detail on the whole material'. In this sense, it was not by accident that we arrived at this example during our general study of the situation of `the one and the many' related to quantum mechanics \cite{aerts1981,aerts1982b,aerts1983}. Of course, there are also many differences between our example about the water vessels and a typical example in quantum mechanics. For example, it is not at all obvious whether the set of states will form a Hilbert space like the set of states of a quantum particle does. On the other hand, we did succeed in putting forward more complicated coupled macroscopic systems that are naturally described by a Hilbert space quantum mechanical model \cite{aerts1991,aertsdhondtdhooghe2005}.

There are two main points we wish to make with respect to our example about the interconnected water vessels and its generalization, and its relation to quantum mechanics. First of all, it becomes much less obvious to imagine a superposition state for a quantum system as a state which is genuinely different from the final state after a measurement, when we consider measurements of position and momentum for a quantum particle. This means that a quantum particle `is really not localized', and `has really no definite momentum' if it is in a state which is a superposition of both position states and momentum states, which is, by the way, the commonest state for a quantum particle. Of course, there is the example of waves, which have, at first sight at least, aspects of these properties. This means that within, let us say, the old paradigm of quantum mechanics, i.e. the wave-particle duality, our proposal to consider superposition states as genuinely different states, would involve giving preference to a wave-like reality for a quantum particle.

However, the second point I want to consider is much more crucial. This is also the point that proves that quantum particles are not waves. Indeed, apart from the fact that in many situations experiments show particle-like behavior in quantum particles, there is a deep structural reason why quantum particles are not waves. Superposition states appear not only as superpositions of, for example, final states after a position measurement of one particle, but also, for example in the case of the presence of two particles, as superpositions of the products of the final states after position measurement of both particles. If we express this in terms of the quantum wave function $\psi(x,y,z)$ of a quantum particle, the quantum wave function of two quantum particles can be said to be a function of the six positions variables $\psi(x_1,y_1,z_1,x_2,y_2,z_2)$, which is in general not the product of two functions of each three positions variables of each of the particles. Going back to our example of the vessels of water, we can find the following exact way of expressing `how a quantum particle is not a wave'. A quantum particle is not an entity `spread out in space', which it would be if it was a wave, but rather an entity `only potentially present in space'. There is a deep and fundamental difference between `spread out in space' and `potentially present in space'. In the case of the vessels, the ten liters of water connected by a tube are potentially divided into $x$ liters and $10-x$ liters, but this volume is not at all `spread out over this range of divisions'. The fact that the division is only potential means that it is not actual, and that it becomes actualized only as a result of the measurement. This is why, in a similar way, a quantum particle in a superposition state `is not inside space', and its `being inside space becomes only actualized due to a position measurement'. And this is also why the wave function of two quantum particles is a function of the six variables, because both particles are not inside space, and, for both of them, the quality of `being inside space' becomes only actualized due to a position measurement', so that $|\psi(x_1,y_1,z_1,x_2,y_2,z_2)|^2$ describes the probability of this happening for particle one at point $(x_1,y_1,z_1)$ `and' for particle two at point $(x_2,y_2,z_2)$. 

It is interesting to note that some recent work, partly inspired by Leggett's inequalities, can be interpreted -- in our opinion at least, we do not have the space in this article to go in depth into this, but will do so in future work -- as a confirmation of this `quantum particle not inside space' view, which we put forward in earlier work for the reasons described above. More specifically, recent attention for `quantum with respect to space' was generated mainly by questions related to the compatibility between relativity and non-locality. In earlier work, we expressed the opinion that one of the main obstacles for a unification between quantum theory and relativity theory consists in a first, not even explicitly stated, axiom of relativity theory, namely that `the collection of all events coincides with a four-dimensional space-time continuum' \cite{aertsaerts2004}. Hence, relativity theory starts off with the prejudice that `reality is contained inside space-time'. Many aspects of our work on quantum mechanics indicate the contrary, namely that reality is not contained inside space-time \cite{aerts1999}. Several authors have paid attention to the problematic relation between quantum mechanics and relativity theory, e.g. with respect to the quantum collapse phenomenon and with respect to the notion of element of reality \cite{aharonovalbert1981,aharonovalbert1984,hardy1992,cohenhiley1995,marchildon2008,marchildon2010}, and more recently an interesting controversy has started based on the so-called Free Will Theorem and explicit models of relativistic collapse theories \cite{goldsteintumulka2003,tumulka2006a,tumulka2006b,conwaykochen2006,tumulka2007,lapiedrasocolovsky2008,conwaykochen2009}.

In our opinion, the very reason for this inconsistency is that the quantum entity itself `is not an entity inside space-time'. For example, the photon pair in a singlet spin state `is not' a pair of photons present in space-time with correlated spins. It is an entity `outside of space-time' and pulled into space-time through the measurements. This explains, for instance, why specific actualizations due to measurements are not co-variant, but only the whole probability distribution is co-variant.

\section{Quantum and Conceptuality}
If quantum particles are not inside space, it means that space is an emergent structure, coming into being jointly with the macroscopic material objects populating it and interacting in it. This interaction consists mainly in bumping and bouncing, and of course also in falling, i.e. moving under the influence of gravity. Could it be that our mathematical model of space has been influenced by the local experience that we as human beings have with these macroscopic material objects that surround us in our habitat on the surface of planet Earth, and whose interaction consists mainly in bumping, bouncing and falling activities? We believe that there is a substantial chance that there is indeed a fundamental influence of this kind. The fact that a stone can be broken down into little pieces, even to the level of grains of sand in the case of sandstone, or into yet smaller particles of dust if brute force is applied and the stone is crushed, has certainly influenced the mathematical model of three-dimensional space -- the theatre of bumping, bouncing and falling performances -- filled with points of zero-size and dimension -- the limits of dust particles of crushed stone.

Let me illustrate this with the following example. People interact with each other in many ways, one of the most important being their communication through language. Language is made up of combinations of words, and words are made up of syllables and encrypted into sounds, in the case of spoken language, or symbols, in the case of written language. It would never occur as a fruitful hypothesis to anybody to consider physical space, i.e. the three-dimensional mathematical space of points, lines, surfaces etc \ldots by which it is described mathematically in physics, as the natural theatre of language. It would never occur as a fruitful hypothesis to anybody either to consider the interaction of people through language as an interaction of the bumping, bouncing and falling type. Although `sound', the carrier of language in the case of spoken language, travels through physical space, this aspect is not at all a fundamental aspect of the interaction pattern connected to language and human beings. This can easily be understood if we think that the encryption of language into sound can be substituted by completely other ways of encryption, e.g. written language, or electromagnetic waves for electronic communication, etc \ldots, without losing its fundamental aspect, and its relation with physical space will change considerably because of it, e.g. written language does not move through space like spoken language does. The fact that language can be carried in so many other ways indicates that language itself does not exist `inside physical space'. It is very well possible, however, that if a language type of interaction exists on the micro-level and a phenomenon is studied on the macro-level, the language interaction that underlies the phenomenon would be hard to notice. In such a case, the micro-interaction pattern would probably be attempted to be explained by means of a bumping and bouncing pattern, while mysterious, unexplicable behavior not fitting the bumping and bouncing pattern would be noticed experimentally. Let us give an example to illustrate this. Suppose that the movement of cars over the surface of planet Earth is studied from a distant star or planet in space, without the alien researchers having any knowledge or suspicion of the underlying micro-interaction of human beings driving these cars. They would notice quite a number of simple regularities. For example, they would identify cities as spots on the surface of planet Earth as attracting large numbers of cars during the morning hours, i.e. when the sun starts to shine, and repelling equally large numbers of cars during the evening hours, when the sun withdraws. If, however, they were to attempt to establish the path of an individual car, they would find that it is not simple at all to do so in the macro-realm, where these attracting and repelling forces exerted by cities on cars exist as a pattern of interaction. Indeed, they would find it easy each morning and also each evening to spot individual cars moving in opposite directions, as well as those that follow completely erratic paths, at times moving with the stream, and at other times suddenly moving in completely different and even opposite directions. If, however, we suppose that the dynamics on the micro-level can be modeled, i.e. the level where human communication, and more generally human thought and decision-making take place, then it would be possible to explain the movements of cars also individually. We believe that the unpredictable nature of the behavior of individual quantum particles might well be due to the existence of a much more complex pattern of interaction on the micro-level than we imagine it to be according to our experience on the macro-level. In \cite{aerts2009a,aerts2010a} we put forward an interpretation for quantum mechanics and an explanatory framework along such lines. More specifically, our recent work on the modeling of human concepts and their combinations, and how they are used in human language, by means of quantum mechanical structures, has inspired us to probe the idea that the interaction regime of quantum particles with matter could be of the language type, i.e. similar to the interaction pattern of language and the human mind.

We believe that it is worth to explore the nature of the `interaction of human minds communicating with each other through language' so as to try to gain a better understanding of quantum mechanics and how it models the behavior of quantum particles interacting with matter. The reason is that we have recently been able to use the quantum mechanical formalism to model mathematically the interaction between human minds through language, which has strongly stimulated the investigation related to this new interpretation for quantum mechanics \cite{aerts2009a,aerts2010a}. The insight that the language interaction regime might be a model for a new interpretation of quantum mechanics has come to us step by step. The first step was related to our quantum-mechanical modeling of a human decision process \cite{aertsaerts1994}. What gave us the idea to model human decision processes by means of quantum mechanics was the insight that quantum mechanics models probability situations where the uncertainty is due not only to a lack of knowledge about a particular existing situation, but possibly also to something else. We thought that as a consequence of this it should be possible to model the situation of a human decision process in which the opinion of the individual deciding is not made up before the decision process starts, i.e. the situation where his or her opinion is partly generated by the unfolding decision process itself.

If we say that people's opinions are not determined prior to the actual start of the decision-making process, this means that the context of decision-making will have an essential impact on the manner in which the decision is ultimately reached. As is commonly known, the quantum formalism is capable of modeling the direct influence of context; there is even a word referring to this aspect of the quantum formalism within quantum mechanics itself, namely `contextuality'. Our next step was therefore to use the quantum formalism to model situations pertaining to domains other than the micro-world where the influence of context is fundamental. This resulted in the elaboration of a theory for the modeling of human concepts based on a generalization of the mathematical formalism of quantum physics, which we called State-Context-Property-Systems (SCOP) \cite{gaboraaerts2002,aertsgabora2005a,aertsgabora2005b}. Context does indeed play a fundamental role in how concepts are combined in human reasoning. If we consider the concept {\it Fruit}, for example, we can experimentally show that {\it Apple} is considered the most typical exemplar of {\it Fruit}. However, if we add the context {\it Tropical} and combine it with {\it Fruit} into {\it Tropical Fruit}, {\it Coconut} proves to be a more typical exemplar than {\it Apple}. More complex combinations give rise to a complicated change of the typicalities of items. The incapacity of existing cognitive science to model these highly contextual effects is generally regarded as one of the major open problems in cognition \cite{hampton1988a,hampton1988b,rips1995}. It is this effect of context which we were able to model by using the quantum formalism and its generalizations \cite{gaboraaerts2002,aertsgabora2005a,aertsgabora2005b}.

However, somewhat unexpectedly, the way in which states of quantum entities are modeled by vectors in a complex Hilbert space also proved to have modeling potential beyond its original application domain, i.e. the micro-world. In \cite{aertsczachor2004}, we showed that the currently most powerful `semantic analysis theories' \cite{saltonwongyang1975}, such as Latent Semantic Analysis (LSA) \cite{deerwesterdumaisfurgaslandauerharshman1990,landauerdumais1977,landauerfoltzlaham1998}, Hyperspace Analogue to Language (HAL) \cite{lundburgess1996}, Probabilistic Latent Semantic Analysis (pLSA) \cite{hofmann1999,vinokourovgirolami2002,gaussiergouttepopatchen2002}, Latent Dirichlet Allocation \cite{bleingjordanlafferty2003} or Topic Model \cite{griffithssteyvers2002}, employ a formal structure containing the quantum formalism as used in our quantum modeling scheme. By the way, this correspondence had already been successfully applied in the field of information retrieval before \cite{widdows2003,widdowspeters2003,vanrijsbergen2004,licunningham2008,zucconazzopardivanrijsbergen2009}. The next step then consisted in showing that the quantum effects of `interference' and `superposition' could model very well so far unexplained and little understood but experimentally well-documented effects. These include problems and fallacies in cognition and decision theory, such as the conjunction fallacy \cite{tverskykahneman1982,tvserskykahneman1983} and the disjunction effect \cite{barhillelneter1986,tverskyshafir1992}. We worked out a quantum-modeling scheme for the application of interference and superposition for the aforementioned type of problems, fallacies and effects \cite{aerts2009b,aertsaertsgabora2009,aertsdhooghe2009}. It is worth noting that the modeling power of quantum structures for concept theories and decision theories has since engendered abundant research activity by a variety of scientists, and that a new research field called `quantum interaction' has emerged from this activity \cite{bruzacole2005,bruzalawlessvanrijsbergensofge2007,bruzalawlessvanrijsbergensofge2008,bruzasofgelawlessvanrijsbergenklusch2009,busemeyerwangtownsend2006,pothosbusemeyer2009,lambertmogilianskyzamirzwirn2009}.  

The formal correspondence, i.e. the quantum formalism as mathematical formalism and modeling tool, has a deeper ground, however. We could show that `superposition' in this quantum-modeling scheme described `the emergence of a new conceptual entity'. In the case of concepts and their combinations, this new conceptual entity is a new concept. Interference models the deviating effect produced by the emergent new concept with respect to a logical analysis of the combination of the original concepts. For example, {\it Olive} scores much higher as a typical exemplar of {\it Fruits or Vegetables} than as a typical exemplar of either {\it Fruits} or {\it Vegetables} based on logical analysis, because it is typical of the new concept {\it `Fruits or Vegetables'}. Contrary to this, {\it Elderberry} scores lower as a typical exemplar of {\it Fruits or Vegetables} than would result from a logical analysis based on its relation with the individual concepts, because {\it Elderberry} is not at all characteristic of the new concept {\it `Fruits or Vegetables'}. We analyzed this modeling of `emergence' by means of `superposition' in great detail in \cite{aerts2009b}, and put it forward in \cite{aertsdhooghe2009} as a basis for a double-layer structure for human thought, i.e. the rational logical versus the intuitive conceptual, often also referred to in dual-process theories \cite{sloman1996,sun2002,baretttugadeengle2004,paivio2007}.

We also found striking structural similarities between our quantum-modeling scheme and approaches developed in artificial intelligence for connectionist models of memory \cite{aertsczachor2004}, more specifically the Holographic Reduced Representations (HRR) of neural networks \cite{plate1995,plate2003}, and we successfully used our approach to this field \cite{aertsczachor2008,aertsczachordemoor2009}.

Of course, the mere fact that the mathematical quantum formalism serves well in modeling concepts and their combinations, and that there are deep structural connections between the quantum mechanical formalism and semantic analysis theories, and connectionist models of memory, does not necessarily mean that the hypothesis of quantum particles themselves being conceptual entities is a fruitful hypothesis. In our opinion, however, there is sufficient reason to make this hypothesis the subject of serious consideration and further elaboration. In \cite{aerts2009a,aerts2010a}, we put forward different aspects that argue in favor of the hypothesis. We will not give a full and systematic discussion of all of these in the present article but highlight some of them and also invoke new arguments to underpin the interpretation worked out in \cite{aerts2009a,aerts2010a}.

\section{Meaning and Coherence}

The human world, as it evolves on the surface of planet Earth, presents us with examples of how conceptual entities, in the form of words, sentences, parts of conversations, written pieces of texts -- i.e. all entities which carry `meaning' -- interact with human minds, with artificial memory structures, with interfaces able to cope with pieces of meaning. This interaction, between entities carrying meaning and interfaces sensitive to meaning, is the essence of human society. It is the dynamics generated by this interaction that drives the evolution of human culture. It is sometimes suggested that humans might well be alone in a large universe of matter and energy, and whenever this idea is expressed, it is usually taken for granted intuitively that in this universe of matter and energy the global interaction pattern is mainly one of the bouncing and bumping type -- and let us repeat that science has proved this intuitive thought to be wrong, since quantum mechanics describing the pattern of interaction in the micro-world demonstrates that the foundations of the interaction pattern of the universe is `not' of the bouncing and bumping type. Sometimes it is imagined that there could be many other life forms in the universe, and that it is only a matter of time before humans will make contact with these other life forms. Such other life forms are generally pictured evolving on the surfaces of planets as well, much like ourselves on our planet Earth. The hypothesis of quantum particles being conceptual entities sheds a different light on the above speculations. It could well be that the pattern of interaction which we see so fruitfully at work in our human society is of a much more general and also much more abundant nature, occurring in many different variations and in many different places and times of the universe. If we consider this from a global perspective, paying due attention to a global Darwinian mechanism, we find that all this is not even unlikely. Indeed, in the first place, we know the pattern exists, because we have the example of human culture on the surface of planet Earth, which means that the laws of nature are compatible with it. We also know that it is an enormously fruitful pattern, with a very strong potential for adaptation and evolution. This is illustrated by the speed of evolution of human culture and its obviously much greater strength and adaptational power than that proposed by biological Darwinian evolution on the surface of planet Earth, which is far more reliant on the bumping and bouncing type of interaction pattern. So, given these elements, it seems rather plausible for evolutions relying on conceptual patterns of interaction to come into existence in other variations in other places and times in the universe. In addition to the overall argument of an evolution process generated within an environment of interaction of the conceptual type being more fecund than an evolution process generated within an environment of interaction of the bumping and bouncing type, we put forward several specific arguments that we think make it more plausible for such a conceptual evolution process to be more abundant than is usually imagined, and to have been taking place on the level of the interaction between quantum particles and matter.

In \cite{aerts2009a} sections 2.2 and 2.3, we showed that with respect to non-locality, more specifically the violation of Bell's inequalities, the role played by `meaning' in the realm of interaction between human minds through language is the equivalent of the role played by `coherence' in the realm of interaction between matter through quantum particles.  In this way, the problem of non-locality gives the impression of somehow being a false problem, because space-time is imposed on the realm of the micro-world as the basic structure, while it is quantum coherence which defines the basic structure of this realm. To give a very concrete example, we can say that, if a photon jumps out of the filament of a shining lamp, it enters a state where it is `not present inside space', so that it would be wrong to picture it traveling from the lamp to someone's eye. Once the photon is captured by the eye, it is sucked into space again, absorbed by the retina of the eye. The foundation of the realm where this phenomenon takes place is structured by coherence, and space-time only plays a secondary role. We can understand that this is possible inside ordinary reality by comparing it with the example of how human minds interact with each other through language. The foundation of the realm where this takes place is structured by meaning, and also here space-time only plays a secondary role. There are indeed many ways in which human minds can exchange meaning, and space-time can play specific roles in each of them, but not the primary role, and it would be wrong to imagine the meaning exchange as primarily a bumping and bouncing type of interaction pattern taking place through space-time, or as `waves of meaning' moving through space in time, with human minds interacting through them.

We also analyzed, in \cite{aerts2009a}, section 4.1, how the duality between the abstract and the concrete in the realm of the human mind and conceptual interaction is the equivalent of the Heisenberg uncertainty in the realm of matter and quantum particles. For example, a concept such as {\it Cat}, which is a very abstract concept, penetrates in a non-local way into many places where clusters of meaning are located at once, and there is no time and space involved because all meaning-producing entities can connect with the concept {\it Cat} at once. Let us put forward again, as we did in \cite{aerts2009a}, the example of the World-Wide Web as a memory structure. Each page of the World-Wide Web which contains the word {\it Cat} has been penetrated by the concept {\it Cat} in this immediate and non-local way. Each page of the World-Wide Web containing the concept {\it Cat} thus produces also a state of the concept {\it Cat}, which is specific to this page. The concept {\it Cat} collapses to this state once it penetrates non-locally into this specific page of the World-Wide Web. And this mechanism of non-local penetration and localization by means of collapse is intrinsic in the meaning-type of interaction which governs human culture. If we allow coherence to be the structuring substance of the micro-realm, a quantum particle in a very non-local state is similar to a very abstract concept, so that it can collapse at any spot of the space-time canvas at once, which indeed accords with the outcome of numerous non-locality experiments carried out in laboratories nowadays.

In \cite{aerts2009a}, section 3, we also explained how a typical interference situation in quantum mechanics can be understood and explained within this conceptual interpretation of quantum mechanics. One of the elements of this explanation is that the localization of a quantum particle in a specific region of space, e.g. the region $A$ behind one of the two slits of an experimental two-slit situation, or the region $B$ between the two slits, should be interpreted in a way that is fundamentally different from the approach that is usually followed in discussing the case where quantum particles are conceptual entities. This localization should not be interpreted as `the quantum particle moving to this region', because this would simply be an instance of approaching the issue within the interaction pattern of a bumping and bouncing type. Indeed, imagining a quantum particle as `moving through space', and hence imagining the process of localization of such a quantum particle as corresponding to `a movement of the quantum particle towards a specific region of space where it then appears in a localized state' are typical examples of thinking in terms of a bumping and bouncing pattern of interaction. If the concept {\it Fruits} became more concrete, for example, one of the possible exemplars of {\it Fruits}, let us say {\it Elderberry}, we would not imagine `{\it Fruits} as moving towards the exemplar region {\it Elderberry}'. We would rather speak of {\it Fruits} becoming more concrete. This insight gives us a way to explain and understand what happens in the double-slit type of situation. If a quantum particle passes through the area of both slits when both are open, it will have a greater tendency subsequently to manifest itself as a quantum particle of which we do not know through which of both slits it has passed, because this is the conceptual content it carries. This means, for example, that if it localizes itself in the area on the screen between both slits, this corresponds well with this conceptual content, because this area makes it clear that it is a quantum particle that raises a lot of doubt as to which the two slits it has passed through. Conversely, such a quantum particle will not easily appear localized right behind either slit, since the spots behind the slits carry the conceptual content of quantum particles that leave no doubt as to the slit they have passed through. And as experiments with the double-slit situation have demonstrated, there is indeed a high probability of detection between both slits and low probabilities of detection right behind both slits. The foregoing is only a brief analysis of the two-slit situation within our new quantum interpretation, and we refer to \cite{aerts2009a} section 3 for a much more detailed description. However, for all its briefness, the analysis given above already makes it clear how all currently known descriptions of the two-slit experiment invariably approach the issue according to a pattern of bumping and bouncing interaction, where the quantum entities are regarded as objects flying through space. It is only when we completely abandon this imagery of bumping and bouncing that we can put forward an explanation we can understand.

One of the aspects of our new interpretation that, to my knowledge, make it stand out from all other existing interpretations of quantum mechanics, is that it also provides an explanation for the behavior of quantum particles with respect to identity and individuality. Again, a detailed account of this aspect of our interpretation can be found in \cite{aerts2009a}. We will only briefly discuss its essence here. Let us consider the conceptual combination {\it Eleven Animals}, and also what this expression means in human language. It indicates eleven entities, each of which is an animal, but in their sense of `animal', they are all perfectly identical. In other words, for human concepts it is possible to understand the complex way in which identity and individuality appear. In fact, there is only one concept {\it Animal} in its very abstract state, but it makes perfect sense to put forward the conceptual combination {\it Eleven Animals}. This creates a different situation, namely `eleven elements, each of which is an animal'. Individuality comes into being only gradually for human concepts. Returning to the conceptual world of the World-Wide Web, we could regard any single webpage as an `individual', since there will only be one webpage that contains exactly the combination of concepts of that particular page. Strictly speaking, even this is not completely true, since it would be possible -- and technically easy -- to make a twin webpage that is completely identical to the first. The reason why the pages of the real World-Wide Web do not have twin pages is because they are meant to be individual entities. They are meant to be the place where conceptual differentiation stops. Each webpage represents a final localized state of all the concepts appearing in it. In our quantum interpretation, physical reality inside space-time is such a final stage of the localization of states of quantum particles, while each quantum particle in its most abstract state, i.e. its state with definite momentum and energy and completely undetermined position and time, is only one. The appearance of `the many' is already linked to a progression of the phenomenon of individualization. We also showed in \cite{aerts2009a} how Bose Einstein statistics naturally follows for the quantum particles in their bosonic states if we carefully analyze the notion of identity and individuality within this interpretation.

We said in \cite{aerts2009a} that we believed the micro realm structured by quantum coherence to be in a much more advanced stage of evolution than is the case for the macro realm structured by meaning. In other words, what surrounds us in the macro realm structured by meaning corresponds to a primitive stage of development, where, for example, the equivalent of space-time has not yet emerged. Let us analyze the way in which our new interpretation sheds light on one of the major problems for all existing interpretations of quantum mechanics, namely the problem related to the question of `why we do not see the superposition states of ordinary macroscopic objects in the everyday world around us'. More concretely, if we consider two possible states of a chair, for example the chair positioned in spot $A$, and the chair positioned in spot $B$, where $A$ and $B$ are regions of space separated from each other, why does a superposition state of these two states not seem to be present in our everyday reality? This superposition state would correspond to a situation in which we would see the chair in $A$ at one moment and in $B$ at another, with the chair always being in the same state and ourselves always looking in the same way. This problem is usually referred to as the Schr\"odinger cat situation. Erwin Schr\"odinger, one of the founding fathers of quantum mechanics, put forward the problem by considering the case of a superposition of a state where a cat is alive and a state where the same cat is dead \cite{schrodinger1935}. To analyze this problem, let us consider again the situation of the World-Wide Web as a memory structure for the macroscopic realm where human concepts are the carriers of meaning. Consider the two logical connectives `or' and `and', which, in the case of concepts, make them change and become either more abstract -- the result of `or' -- or more concrete -- the result of `and'. If we consider two webpages, `webpage $A$' and `webpage $B$', we can say that whereas `webpage $A$ or webpage $B$' is `not' a webpage, `webpage $A$ and webpage $B$' can be made into a webpage. While we may bring all the text of webpage $A$ and that of webpage $B$ together under one url and call it `webpage $A$ and webpage $B$', we could not do the same with `webpage $A$ or webpage $B$'. It is not possible to attribute an url to a situation corresponding to `webpage $A$ or webpage $B$'. The reason for this is that webpages have acquired the status of objects in our human world. And indeed, if we consider two objects, $A$ and $B$, then `$A$ or $B$' is `not' an object, whereas `$A$ and $B$' is. For concepts, however, the symmetry between the `or' and the `and' connectives has remained intact. Indeed, if we consider two concepts, $A$ and $B$, then `$A$ or $B$' is again a concept, and also `$A$ and $B$' is again a concept. So, within the new interpretation we put forward, the fundamental question with respect to the Schr\"odinger's cat situation is: `Why have macroscopic constellations of quantum particles acquired the status of objects?'.

We want to put forward a possible explanation for the Schr\"odinger cat problem within our new quantum interpretation. However, we should add that, in using the example of the World-Wide Web as a conceptual environment to gain a better insight into the Schr\"odinger cat problem, even if we accept the new interpretation put forward in \cite{aerts2009a,aerts2010a}, we must acknowledge that the explanation we put forward is speculative. A thorough investigation of the possible ways for space-time to emerge from a more primitive situation similar to that of human concepts and their environment could yield additional possibilities to those we put forward here. The reason why we believe our discussion is valuable despite this speculative nature is that all attempts to explain the Schr\"odinger cat problem have so far been very speculative anyway.

So the question is: `Why have webpages acquired the status of objects?'. We believe that we should rather speak of a process of objectification. In our opinion, neither webpages nor material entities, i.e. constellations of quantum particles, are really objects. Although they have evolved towards a type of concept that is closer to an object, they remain concepts. Of course, we should add to this that it is an essential aspect of a concept to reveal its conceptual nature only in the presence of an interface that is capable of interpreting - and hence reacting with Ð it conceptually. If we say that material entities like chairs have undergone a process of objectification, it also means that there are no interfaces available to react with them in a conceptual way and reveal their conceptual nature. The following example aims to provide a better insight into this. We consider the concepts {\it Fruits}, {\it Vegetables}, {\it Furniture} and {\it Bird} and investigate their presence in the environment of the World-Wide Web. Using Google on April 25, 2010, we found that there were 55,400,000 hits of {\it Fruits}, 42,600,000 hits of {\it Vegetables}, 184,000,000 hits of {\it Furniture} and 291,000,000 hits of {\it Bird}. In other words, the four concepts appear in comparably large numbers. There are 1,400,000 hits of `{\it Fruits or Vegetables}' and only 34,000 hits of `{\it Furniture or Bird}'. What is the reason for this big difference? Since we use our human mind to determine the meaning content of the webpages, the answer is easy to give. There are many more sentences, pieces of text, etc. \ldots containing the combination of concepts {\it Fruits or Vegetables} than sentences, pieces of text, etc. \ldots containing the combination {\it Bird or Furniture}. The reason is that in the world we live in, {\it Fruits or Vegetables} is a more meaningful combination of concepts than {\it Bird or Furniture}. Let us make this difference even more pronounced. If we enter the two concepts {\it Fruits} and {\it Vegetables} in Google, the first meaningful sentence we see on a webpage is the following: `Eat a colorful variety of fruits and vegetables every day for better health'. Suppose we take a combination of four concepts from this sentence, e.g. `colorful variety of fruits and vegetables', and we look up the number of webpages that contain this exact combination of four concepts, we find there are 108,000 of such pages. If we do the same for the two concepts {\it Furniture} and {\it Bird}, the first meaningful piece of sentence of four words containing the two concepts is `Indonesian furniture bird cage'. If we look up the number of webpages containing this exact combination of concepts, we find that there is exactly one such page, viz. \url{http://www.indonesia-furnitures.com/page_thumbnail/birdcages_thumbnail.html}. 

This analysis shows that, in the case of human concepts, some pairs of concepts are much more submerged in each other's `meaning range' than other pairs of concepts. It is very well possible that the clustering of concepts, which is a consequence of this phenomenon, is inherent in the way a `meaning type of interaction' works. If coherence is the equivalent of meaning in the case of quantum particles, then plausibly also coherence brings about this type of clustering. And indeed, the notion of `coherence length', as a measure attributable to a specific quantum state and indicating the range in which it might interact coherently with another quantum state, reveals the presence of a similar type of clustering. The clustering brings about an objectification process, larger clusters attaining a stronger object status within the governing coherence-meaning type of interaction. Space-time, as a `container of objects', has emerged jointly with this clustering and objectification process for the clusters. Superpositions between clusters are almost absent. These clusters of quantum particles, still essentially concepts in nature, have evolved towards an object-like status, and hence in our first physical theories, called classical physics, we have started to model them as objects. However, their essential conceptual nature can still be revealed if appropriate experimental conditions for this are realized. And indeed, microscopic quantum effects have been shown to be capable of penetrating the macroscopic realm in experiments with Bose Einstein condensates \cite{andersonenshermatthewswiemancornell1995,davismewesandrewsvandrutendurfeekurnketterle1995}, the phenomena of Macroscopic Quantum Coherence and Macroscopic Quantum Tunneling in SQUIDS \cite{cosmellicarelliacastellanochiarelloleoniatorriolia2002,coratorombettosilvestrinigranatarussoruggiero2004}, revealing superposition states of biomolecules and fluorofullerenes \cite{hackermulleruttenthalerhornbergerreigerbrezgerzeilingerarndt2003}, but also in recent findings of quantum effects in biological systems, for example with respect to the mechanism of photosynthesis \cite{engelcalhounreadahnmancalchengblankenshipfleming2007,scholes2010}, and of macroscopic electronic circuits behaving quantum mechanically \cite{ansmannwangbialczakhofheinzluceroneeleyoconnellsankweideswennerclelandmartinis2009}, or classical mechanical systems controlled by quantum mechanics \cite{connellhofheinzansmannbialczaklenanderluceroneeleysankwangweideswennermartiniscleland}. However, for these effects to be realized, very carefully designed laboratory situations need to be prepared, while the quantum biological effects probably required a long period of evolution. This means that common objects in the world around us do not appear in quantum superposition states because these states have not been selected within the coherence type of interaction. If we take into account that the conceptual nature of an entity depends equally so on the interface being able to conceptually interact with this entity, it is as correct to say that common objects do not appear in quantum superposition states, because there is no interface capable of interacting with them. There is, by the way, yet another way to look at this situation and one that is worth mentioning, because, although compatible with the foregoing, it sheds new light on the matter. It is possible to interpret the situation in such a way that the superposition state of, for example, two chairs in different locations $A$ and $B$, exists without being a state of a concept which can be interpreted as an object. And there is an interface capable of interacting with this superposition state, namely the human mind. What we mean is that this superposition state is the concept `The chair in spot $A$ or in spot $B$'. This is not an object but a concept, and our human mind can conceptually interact with it and hence act as an interface for it. Indeed, this is exactly what happens when we are in a situation where we do not know whether the chair in question is in spot $A$ or in spot $B$. If we prefer to interpret the situation in this manner, it means that we see objects as states of concepts collapsed in space-time -- what psychologists call instantiations of concepts. Superposition is then an operation which preserves neither the instantiation quality of a concept nor the localization quality of a quantum particle state.


\begin{thebibliography}{99}

\bibitem{aerts2009a} Aerts, D. (2009). Quantum particles as conceptual entities. A possible explanatory framework for quantum theory. {\it Foundations of Science}, {\bf 14}, pp. 361-411.

\bibitem{aerts2010a} Aerts, D. (2010). Interpreting quantum particles as conceptual entities. {\it International Journal of Theoretical Physics}.

\bibitem{copernicus1543} Copernicus, N. (1543). {\it De Revolutionibus Orbium Coelestium}. Norimbergae: J. Petreium.

\bibitem{mohapatra2003} Mohapatra, R. N. (2003). {\it Unification and Supersymmetry: The Frontiers of Quark-Lepton Physics}. New York: Springer.

\bibitem{schumm2004} Schumm, B. A. (2004). {\it Deep Down Things: The Breathtaking Beauty of Particle Physics}. Baltimore, Maryland: The Johns Hopkins University Press.

\bibitem{smolin2006} Smolin, L. The Trouble With Physics: The Rise of String Theory, The Fall of a Science, and What Comes Next (2006). New York: Mariner Books.

\bibitem{wilczek2008} Wilczek, F. (2008). {\it Lightness of Being: Mass, Ether, and the Unification of Forces}. New York: Basic Books.

\bibitem{einsteinpodolskyrosen1935} Einstein, A., Podolsky, B. and Rosen, N. (1935). Can quantum-mechanical description of physical reality be considered complete? {\it Physical Review}, {\bf 47}, pp. 777-780.

\bibitem{bohm1952} Bohm, D. (1952). A suggested interpretation of the quantum theory in terms of `hidden' variables I \& II. {\it Physical Review}, {\bf 85}, pp. 166Ð179; pp. 180Ð193.

\bibitem{bell1964} Bell, J. S. (1964). On the Einstein Podolsky Rosen Paradox. {\it Physics}, {\bf 1}, pp. 195-200.

\bibitem{bell1966} Bell, J.S. (1966). On the problem of hidden variables in quantum mechanics. {\it Reviews of Modern Physics}, {\bf 38}, pp. 447Ð452.

\bibitem{clauserhorne1974} Clauser, J.F. and Horne, M.A. (1974). Experimental consequences of objective local theories. {\it Physical Review D}, {\bf 10}, pp. 526-535.

\bibitem{clausershimony1978} Clauser, J. F. and Shimony, A. (1978). Bell's theorem: experimental tests and implications. {\it Reports on Progress in Physics}, {\bf 41}, pp. 1881-1927.

\bibitem{aspect1981} Aspect, A., Grangier, P. and Roger, G. (1981). Experimental tests of realistic local theories via Bell's theorem. {\it Physical Review Letters}, {\bf 47}, pp. 460-463.

\bibitem{bell1987} Bell, J.S. (1987). {\it Speakable and Unspeakable in Quantum Mechanics}. Cambridge: Cambridge University Press. 

\bibitem{aerts1982a} Aerts, D. (1982). Example of a macroscopical situation that violates Bell inequalities. {\it Lettere al Nuovo Cimento}, {\bf 34}, pp. 107-111.

\bibitem{aerts1985} Aerts, D. (1985). The physical origin of the EPR paradox and how to violate Bell inequalities by macroscopical systems. In P. Lathi and P. Mittelstaedt (Eds.), {\it Symposium on the Foundations of Modern Physics: 50 years of the Einstein-Podolsky-Rosen Gedankenexperiment} (pp. 305-320). Singapore: World Scientific.

\bibitem{aerts1991} Aerts, D. (1991). A mechanistic classical laboratory situation violating the Bell inequalities with 2$\sqrt(2)$, exactly 'in the same way' as its violations by the EPR experiments. {\it Helvetica Physica Acta}, {\bf 64}, pp. 1-23.

\bibitem{aertsaertsbroekaertgabora2000} Aerts, D., Aerts, S., Broekaert, J. and Gabora, L. (2000). The violation of Bell inequalities in the macroworld. {\it Foundations of Physics}, {\bf 30}, pp. 1387-1414.

\bibitem{aertsczachordhooghe2006} Aerts, D., Czachor, M. and D'Hooghe, B. (2006). Towards a quantum evolutionary scheme: violating Bell's inequalities in language. In N. Gontier, J. P. Van Bendegem and D. Aerts (Eds.), {\it Evolutionary Epistemology, Language and Culture - A non adaptationist systems theoretical approach}. Dordrecht: Springer.

\bibitem{leggett2003} Leggett, A. J. (2003). Nonlocal hidden-variable theories and quantum mechanics: An incompatibility theorem. {\it Foundations of Physics}, {\bf 33}, pp. 1469-1493.

\bibitem{groblacherpaterekkaltenbaekbruknerzukowskiaspelmeyerzeilinger2007} Gr\"oblacher, S., Paterek, T., Kaltenbaek, R., Brukner, C., Zukowski, M., Aspelmeyer, M. and Zeilinger, A. (2007). An experimental test of non-local realism. {\it Nature}, {\bf 446}, pp. 871-875.

\bibitem{paterekfedrizzigroblacherjenneweinzukowskiaspelmeyerzeilinger2007} Paterek, T., Fedrizzi, A., Gr\"oblacher, S., Jennewein, Th., Zukowski, M., Aspelmeyer, M. and Zeilinger A. (2007). Experimental test of non-local realistic theories without the rotational symmetry assumption. {\it Physical Review Letters}, {\bf 99}, 210406.

\bibitem{branciardlinggisinkurtsieferlamaslinaresscarani2007} Branciard, C., Ling, A., Gisin, N., Kurtsiefer, Ch., Lamas-Linares, A. and Scarani, V. (2007). Experimental falsification of Leggetts non-local variable model. {\it Physical Review Letters}, {\bf 99}, 210407.

\bibitem{brianciardbrunnergisinkurtsieferlamaslinareslingscarani2008}Branciard, C., Brunner, N. Gisin,  N. Kurtsiefer,  C. Lamas-Linares, A., Ling, A. and Scarani, V. (2008). Testing quantum correlations versus single-particle properties within Leggett's model and beyond. {\it Nature Physics}, {\bf 4}, pp. 681-685.

\bibitem{eisamangoldschmidtchenfanmigdall2008} Eisaman, M. D., Goldschmidt, E. A., Chen, J., Fan, J. and Migdall, A. (2008). Experimental test of nonlocal realism using a fiber-based source of polarization-entangled photon pairs. {\it Physical Review A}, {\bf 77}, 032339.

\bibitem{aerts1981} Aerts, D. (1981). {\it The One and the Many: Towards a Unification of the Quantum and Classical Description of One and Many Physical Entities}. Doctoral dissertation, Brussels Free University.

\bibitem{aerts1982b} Aerts, D. (1982) Description of many physical entities without the paradoxes encountered in quantum mechanics. {\it Foundations of Physics}, {\bf 12}, pp. 1131-1170.

\bibitem{aerts1983} Aerts, D. (1983). Classical-theories and non-classical theories as a special case of a more general theory. {\it Journal of Mathematical Physics}, {\bf 24}, pp. 2441-2453.

\bibitem{aertsdhondtdhooghe2005} Aerts, D., D'Hondt, E. and D'Hooghe, B. (2005). A geometrical representation of entanglement as internal constraint. {\it International Journal of Theoretical Physics}, {\bf 44}, pp. 897-907.

\bibitem{aertsaerts2004} Aerts, D. and Aerts. S. (2004). Towards a general operational and realistic framework for quantum mechanics and relativity theory. In A. C. Elitzur, S. Dolev and N. Kolenda (Eds.), {\it Quo Vadis Quantum Mechanics? Possible Developments in Quantum Theory in the 21st Century} (pp. 153-208). New York: Springer.

\bibitem{aerts1999} Aerts, D. (1999). The stuff the world is made of: physics and reality. In D. Aerts, J. Broekaert and E. Mathijs (Eds.), {\it Einstein meets Magritte: An Interdisciplinary Reflection} (pp. 129-183). Dordrecht: Kluwer Academic.

\bibitem{aharonovalbert1981} Aharonov, Y. and Albert, D. Z. (1981). Can we make sense out of the measurement process in relativistic quantum mechanics? {\it Physical Review D}, {\bf 24}, pp. 359-370. 

\bibitem{aharonovalbert1984} Aharonov, Y. and Albert, D. Z. (1984). Is the usual notion of time evolution adequate for quantum mechanical systems? {\it Physical Review D}, {\bf 29}, pp. 223-227.

\bibitem{hardy1992} Hardy, L. (1992). Quantum mechanics, local realistic theories, and Lorentz-invariant realistic theories. {\it Physical Review Letters}, {\bf 68}, pp. 2981-2984.

\bibitem{cohenhiley1995} Cohen, O. and Hiley, B. J. (1995). Reexamining the assumption that elements of reality can be Lorentz invariant. {\it Physical Review A}, {\bf 52}, pp. 76-81.

\bibitem{marchildon2008} Marchildon, L. (2008). On relativistic elements of reality. {\it Foundations of Physics}, {\bf 38}, pp. 804-817.

\bibitem{marchildon2010} Marchildon, L. (2010). Hardy's setup and elements of reality. {\it Physica E: Low-dimensional Systems and Nanostructures}, {\bf 42}, pp. 323-326.

\bibitem{goldsteintumulka2003} Goldstein, S., Tumulka, R. (2003). Opposite arrows of time can reconcile relativity and
nonlocality. {\it Classical and Quantum Gravity}, {\bf 20}, pp. 557Ð564.

\bibitem{tumulka2006a} Tumulka, R. (2006). A relativistic version of the Ghirardi-Rimini-Weber model. {\it Journal of Statistical Physics}, {\bf 125}, pp. 821-840.

\bibitem{tumulka2006b} Tumulka, R. (2006). Collapse and relativity. Archive reference and link \url{http://uk.arxiv.org/abs/quant-ph/0602208}

\bibitem{conwaykochen2006} Conway, J.H. and Kochen, S. (2006). The Free Will Theorem. {\it Foundations of Physics}, {\bf 36}, pp. 1441-1473.

\bibitem{tumulka2007} Tumulka, R. (2007). Comment on `The Free Will Theorem'. {\it Foundations of Physics}, {\bf 37}, pp. 186-197.

\bibitem{lapiedrasocolovsky2008} Lapiedra, R. and Socolovsky, M. (2008). Leggett inequalities and the completeness of quantum mechanics. Archive reference and link \url{http://arxiv.org/pdf/0806.2037}

\bibitem{conwaykochen2009} Conway, J.H. and Kochen, S. (2009). The Strong Free Will Theorem. {\it Notices of the American Mathematical Society}, {\bf 56}, 226.

\bibitem{aertsaerts1994} Aerts, D., Aerts, S. (1994). Applications of quantum statistics in psychological studies of decision processes. {\it Foundations of Science}, {\bf 1}, pp. 85-97.

\bibitem{gaboraaerts2002} Gabora, L. and Aerts, D. (2002). Contextualizing concepts using a mathematical generalization of the quantum formalism. {\it Journal of Experimental and Theoretical Artificial Intelligence}, {\bf 14}, pp. 327-358. 

\bibitem{aertsgabora2005a} Aerts, D. and Gabora, L. (2005). A theory of concepts and their combinations I: The structure of the sets of contexts and properties. {\it Kybernetes}, {\bf 34}, pp. 167-191.

\bibitem{aertsgabora2005b} Aerts, D. and Gabora, L. (2005). A theory of concepts and their combinations II: A Hilbert space representation. {\it Kybernetes}, {\bf 34}, pp. 192-221.

\bibitem{hampton1988a} Hampton, J. A. (1988). Disjunction of natural concepts. {\it Memory \& Cognition}, {\bf 16}, pp. 579-591. 

\bibitem{hampton1988b} Hampton, J. A. (1988). Overextension of conjunctive concepts: Evidence for a unitary model for concept typicality and class inclusion. {\it Journal of Experimental Psychology: Learning, Memory, and Cognition}, {\bf 14}, pp. 12-32.

\bibitem{rips1995} Rips, L. J. (1995). The current status of research on concept combination. {\it Mind \& Language}, {\bf 10}, pp. 72-104.

\bibitem{aertsczachor2004} Aerts, D. and Czachor, M. (2004). Quantum aspects of semantic analysis and symbolic artificial intelligence. {\it Journal of Physics A-Mathematical and General}, {\bf 37}, pp. L123-L132.

\bibitem{saltonwongyang1975} Salton, G., Wong, A. and Yang, C. S. (1975). A vector space model for automatic indexing. {\it Communications of the ACM}, {\bf 18}, pp. 613-620.

\bibitem{deerwesterdumaisfurgaslandauerharshman1990} Deerwester, S., Dumais, S. T., Furnas, G. W., Landauer, T. K. and Harshman, R. (1990). Indexing by Latent Semantic Analysis. {\it Journal of the American Society for
Information Science}, {\bf 41}, pp. 391-407.

\bibitem{landauerdumais1977} Landauer, T. K. and Dumais, S. T. (1977). Solution to Plato's Problem: The Latent Semantic Analysis theory of acquisition, induction and representation of knowledge. {\it Psychological Review}, {\bf 104}, pp. 211-240.

\bibitem{landauerfoltzlaham1998} Landauer, T. K., Foltz, P. W and Laham, D. (1998). Introduction to Latent Semantic Analysis. {\it Discourse Processes}, {\bf 25}, pp. 259-284.

\bibitem{lundburgess1996} Lund, K and Burgess, C. (1990). Producing high-dimensional semantic spaces from lexical co-occurence. {\it Behavior Research Methods, Instruments and Computers}, {\bf 28}, pp. 203-208.

\bibitem{hofmann1999} Hofmann, T. (1999). Probabilistic Latent Semantic Analysis. {\it Proceedings of Uncertainty in Artificial Intelligence, UAI'99}. Stockholm.

\bibitem{vinokourovgirolami2002} Vinokourov, A. and Girolami, M. (2002). A probabilistic framework for the hierarchic organisation and classification of document collections. {\it Journal of Intelligent Information Systems}, {\bf 18}, pp. 153-172.

\bibitem{gaussiergouttepopatchen2002} Gaussier, E., Goutte, C., Popat, K. and Chen, F. (2002). A hierarchical model for clustering and categorising documents. {\it Advances in Information Retrieval, Lecture Notes in Computer Science}, {\bf 2291}, pp. 121-125.

\bibitem{bleingjordanlafferty2003} Blei, D. M., Ng, A. N., Jordan, M. I. and Lafferty, J. (2003). Latent Dirichlet Allocation. {\it Journal of Machine Learning Research}, {\bf 3}, pp. 993-1022.

\bibitem{griffithssteyvers2002} Griffiths, T. L. and Steyvers, M. (2002). Prediction and semantic association. {\it Advances in Neural Information Processing Systems}.
   
\bibitem{widdows2003} Widdows, D. (2003). Orthogonal negation in vector spaces for modeling word-meanings and document retrieval. In {\it Proceedings of the 41st Annual Meeting of the
Association for Computational Linguistics} (pp. 136-143). Sapporo, Japan, July 7-12.

\bibitem{widdowspeters2003} Widdows, D. and Peters, S. (2003). Word vectors and quantum logic: Experiments with negation and disjunction. In {\it Mathematics of Language  8} (pp. 141-154).
Indiana: Bloomington.

\bibitem{vanrijsbergen2004} Van Rijsbergen, K. (2004). {\it The Geometry of Information Retrieval}. Cambridge UK: Cambridge University Press.

\bibitem{licunningham2008} Li, Y. and Cunningham, H. (2008). Geometric and quantum methods for information retrieval. {\it ACM SIGIR}, {\bf 42}, pp. 22-32.

\bibitem{zucconazzopardivanrijsbergen2009} Zuccon, G., Azzopardi, L and van Rijsbergen, K. (2009). The quantum probability ranking principle for information retrieval. {\it Lecture Notes in Computer Science}, 
{\bf 5766}, pp. 232-240. Berlin: Springer.

\bibitem{tverskykahneman1982} Tversky, A. and Kahneman, D. (1982). Judgments of and by representativeness. In D. Kahneman, P. Slovic and A. Tversky (Eds.), {\it Judgment Under Uncertainty: Heuristics and Biases}. Cambridge, UK: Cambridge University Press.

\bibitem{tvserskykahneman1983} Tversky, A. and Kahneman, D. (1983). Extensional versus intuitive reasoning: The conjunction fallacy in probability judgment. {\it Psychological Review}, {\bf 90}, pp. 293-315.

\bibitem{barhillelneter1986} Bar-Hillel, M. and Neter, E. (1986). How alike is it? versus how likely is it?: A disjunction fallacy in probability judgments. {\it Journal of Personality and Social Psychology}, {\bf 65}, pp. 1119-1131.

\bibitem{tverskyshafir1992} Tversky, A. and Shafir, E. (1992). The disjunction effect in choice under uncertainty. {\it Psychological Science}, {\bf 3}, pp. 305Ð309.

\bibitem{aerts2009b} Aerts, D. (2009). Quantum structure in cognition. {\it Journal of Mathematical Psychology}, {\bf 53}, pp. 314-348.

\bibitem{aertsaertsgabora2009} Aerts, D., Aerts, S. and Gabora, L. (2009). Experimental evidence for quantum structure in cognition. In Proceedings of QI 2009-Third International Symposium on
Quantum Interaction, Lecture Notes in Computere Science, pp. 59Ð70. Berlin: Springer.

\bibitem{aertsdhooghe2009} Aerts, D. and D'Hooghe, B. (2009). Classical logical versus quantum conceptual thought: Examples in economics, decision theory and concept theory. In Proceedings
of QI 2009-Third International Symposium on Quantum Interaction, Lecture Notes in Computere Science (pp. 128Ð142). Berlin: Springer.

\bibitem{bruzacole2005} Bruza, P. D. and Cole, R. J. (2005). Quantum logic of semantic space: An exploratory investigation of context effects in practical reasoning. In S. Artemov, H.
Barringer, A. S. d'Avila Garcez, L. C. Lamb and J. Woods (Eds.), {\it We will show them: Essays in honour of Dov Gabbay}. London: College Publications.

\bibitem{bruzalawlessvanrijsbergensofge2007} Bruza, P. D., Lawless, W., van Rijsbergen, C. J. and Sofge, D. (Eds.), (2007). {\it Proceedings of the AAAI spring symposium on quantum interaction}. California: AAAI
Press. 

\bibitem{bruzalawlessvanrijsbergensofge2008} Bruza, P. D., Lawless, W., van Rijsbergen, C. J. and Sofge, D. (Eds.), (2008). {\it Quantum interaction: Proceedings of the second quantum interaction symposium}. London: College Publications. 

\bibitem{bruzasofgelawlessvanrijsbergenklusch2009} Bruza, P. D., Sofge, D., Lawless, W., van Rijsbergen, C. J. and Klusch, M. (Eds.), (2009). Proceedings of the third quantum interaction symposium. Lecture notes in
artificial intelligence 5494. Berlin, Heidelberg: Springer.

\bibitem{busemeyerwangtownsend2006} Busemeyer, J. R., Wang, Z. and Townsend, J. T. (2006). Quantum dynamics of human decision making. {\it Journal of Mathematical Psychology}, {\bf 50}, pp. 220-241.

\bibitem{pothosbusemeyer2009} Pothos, E. M. and Busemeyer, J. R. (2009). A quantum probability model explanation for violations of `rational' decision theory. {\it Proceedings of the Royal Society, B}.

\bibitem{lambertmogilianskyzamirzwirn2009} Lambert Mogiliansky, A., Zamir, S and Zwirn, H. (2009). Type indeterminacy: A model of the KT(KahnemanÐTversky)-man. {\it Journal of Mathematical Psychology}, {\bf 53},
pp. 349-361.

\bibitem{sloman1996} Sloman S.A. (1996). The empirical case for two systems of reasoning. {\it Psychological Bulletin}, {\bf 119}, pp. 3-22.

\bibitem{sun2002} Sun, R. (2002). {\it Duality of the Mind}. Mahwah, NJ: Lawrence Erlbaum Associates.

\bibitem{baretttugadeengle2004} Barrett, L. F., Tugade, M. M. and Engle, R. W. (2004) Individual differences in working memory capacity and dual-process theories of the mind. {\it Psychological Bulletin}, {\bf 130}, pp. 553-573.

\bibitem{paivio2007} Paivio, A. (2007). {\it Mind and Its Evolution: A Dual Coding Theoretical Approach}. Mahwah, NJ. Lawrence Erlbaum Associates.

\bibitem{plate1995} Plate, T. A. (1995). Holographic reduced representations. {\it IEEE Transactions on Neural Networks}, {\bf 6}, pp. 623-641. 

\bibitem{plate2003} Plate, T. A. (2003). {\it Holographic Reduced Representation: Distributed Representation for Cognitive Structures}. CSLI-publications: University of Chicago Press.

\bibitem{aertsczachor2008} Aerts, D. and Czachor, M. (2008). Tensor-product vs. geometric-product coding. {\it Physical Review A}, {\bf 77}, 012316. 

\bibitem{aertsczachordemoor2009} Aerts, D., Czachor, M. and De Moor, B. (2009). Geometric analogue of holographic reduced representation. {\it Journal
of Mathematical Psychology}, {\bf 53}, pp. 389-398.

\bibitem{schrodinger1935} Schr\"odinger, E. (1935). Die gegenw\"artige Situation in der Quantenmechanik. {\it Naturwissenschaften}, {\bf 23}, pp. 823-828.

\bibitem{andersonenshermatthewswiemancornell1995} Anderson, M. H., Ensher, J. R., Matthews, M. R., Wieman, C. E. and Cornell, E. A. (1995). Observation of Bose-Einstein condensation in a dilute atomic vapor. {\it Science}, {\bf 269}, pp. 198-201.

\bibitem{davismewesandrewsvandrutendurfeekurnketterle1995} Davis, K.B., Mewes, M.O., Andrews, M.R., van Druten, N.J., Durfee, D.S., Kurn, D.M. and Ketterle, W. (1995). Bose-Einstein condensation in a gas of sodium atoms. {\it Physical Review Letters}, {\bf 75}, pp. 3969-3973.

\bibitem{cosmellicarelliacastellanochiarelloleoniatorriolia2002} Cosmelli, C., Carellia, P., Castellano, M. G., Chiarello, F., Leonia, R. and Torriolia, G. (2002). Measurements for an experiment of macroscopic quantum coherence with SQUIDs. {\it Physica C : Superconductivity},
{\bf 372-376} Part 1, pp. 213-216.

\bibitem{coratorombettosilvestrinigranatarussoruggiero2004} Corato, V., Rombetto, S., Silvestrini, P., Granata, C., Russo, R. and Ruggiero, B. (2004). Observation of macroscopic quantum tunnelling in a rf superconducting quantum interference device system. {\it Superconductor Science and Technology}, {\bf 17}, S385.

\bibitem{hackermulleruttenthalerhornbergerreigerbrezgerzeilingerarndt2003} Hackerm\"uller, L., Uttenthaler, S., Hornberger,  K., Reiger, E., Brezger, B., Zeilinger, A. and Arndt, M. (2003). Wave nature of biomolecules and fluorofullerenes. {\it Physical Review Letters}, {\bf 91}, 090408. 

\bibitem{engelcalhounreadahnmancalchengblankenshipfleming2007} Engel, G.S., Calhoun, T.R., Read, E.L., Ahn,  T.K., Mancal, T., Cheng, Y-C., Blankenship, R.E. and Fleming, G.R. (2007). Evidence for wavelike energy transfer through quantum coherence in photosynthetic complexes. {\it Nature}, {\bf 446}, pp. 782-786.

\bibitem{scholes2010} Scholes, G. D. (2010). Quantum-coherent electronic energy transfer: Did Nature think of it first? {\it Journal of Physical Chemistry Letters}, {\bf 1}, pp. 2Ð8.

\bibitem{ansmannwangbialczakhofheinzluceroneeleyoconnellsankweideswennerclelandmartinis2009} Ansmann, M., Wang, H., Bialczak, R. C., Hofheinz, M., Lucero, E. Neeley, M., O'Connell, A. D., Sank, D., Weides,  M., Wenner, J., Cleland, A. N. and Martinis, J. M. (2009).  Violation of Bell's inequality in Josephson phase qubits. {\it Nature}, {\bf 461}, pp. 504-506.  

\bibitem{connellhofheinzansmannbialczaklenanderluceroneeleysankwangweideswennermartiniscleland} O'Connell, A. D., Hofheinz, M., Ansmann, M., Bialczak, R. C., Lenander, M., Lucero, E., Neeley, M., Sank, D., Wang, H., Weides1, M., Wenner, J., Martinis, J. M. \& Cleland, A. N. (2010). Quantum ground state and single-phonon control of a mechanical resonator. {\it Nature},  {\bf 464}, pp. 697-703.

\end{thebibliography}
\end{document}